\documentclass[twocolumn,aps,pra,showpacs,groupedaddress]{revtex4-1}
\usepackage[matrix,frame,arrow]{xy}
\usepackage{amsmath}

\usepackage{epsfig,amssymb,amsmath,mathrsfs,color} 
\usepackage[active]{srcltx} 
\usepackage[hypertex,linkcolor=red]{hyperref}
\usepackage[matrix,frame,arrow]{xypic}
\usepackage{tikz}
 \newtheorem{definition}{Definition}

 \def\>{\rangle}
\def\<{\langle}

\newcommand{\figvi}{\begin{center}
\begin{tikzpicture}
\begin{scope}[scale=1.2]
\draw[color=black,ultra thick] (2.25,1.75)--(1.75,1.25)--(1.25,1.75)--(1.75,2.25)--(2.25,1.75)--(1.75,1.25);
\draw[color=black,ultra thick]
(-0.35,-1.5)--(0.35,-1.5)--(0.35,-2.2)--(-0.35,-2.2)--(-0.35, -1.5);
\draw (0,-1.85) node {$c_{k_0}$};
\draw[color=black,ultra thick] (-2.25,1.75)--(-1.75,1.25)--(-1.25,1.75)--(-1.75,2.25)--(-2.25,1.75)--(-1.75,1.25);
\draw (-1.75,1.8) node {$a_{i_0}$};
\draw[color=black,ultra thick, ->] (0,0)--(1.5,1.5); 
\draw[color=black,ultra thick, ->] (0,0)--(-1.5,1.5); 
\draw[color=black,ultra thick, ->] (0,-1.5)--(0,0);
\draw (-1, -1.8) node {{$D_C$}};
\draw (0.,-3.) node {{\footnotesize {\bf Figure 6}: Scheme for a generic
    quantum experiment involving}};
\draw (-0.18, -3.25) node {\footnotesize three sets of random
  outcomes. The outcome on device D$_C$};
\draw (-1.45, -3.5) node {\footnotesize causes those on devices D$_A$ and D$_B$.};
\draw (1.75,1.8) node {{$b_{j_0}$}};
\draw (2.3,1) node {{ $D_B$ }}; 
\draw (-2.3, 1) node {{ $D_A$ }};
\draw (0.75,1.1) node  [rotate=45] {{ $\mathscr{S}_B$ }};
\draw (0.2,-0.8) node [rotate = 90] {$\mathscr{S}_C$};
\draw (-0.75, 1.1) node [rotate = 315] {{ $\mathscr{S}_A$ }};
\draw (0.3,0.) node {{$\mathcal{T}$}};
\end{scope}
\end{tikzpicture}
\end{center}}

\newcommand{\figv}{\begin{center}
\begin{tikzpicture}
\begin{scope}[scale=1.2]
\draw[color=black,ultra thick] (2.25,1.75)--(1.75,1.25)--(1.25,1.75)--(1.75,2.25)--(2.25,1.75)--(1.75,1.25);
\draw[color=black,ultra thick]
(-0.35,-1.5)--(0.35,-1.5)--(0.35,-2.2)--(-0.35,-2.2)--(-0.35, -1.5);
\draw (0,-1.85) node {$c_{k_0}$};
\draw[color=black,ultra thick] (-2.25,1.75)--(-1.75,1.25)--(-1.25,1.75)--(-1.75,2.25)--(-2.25,1.75)--(-1.75,1.25);
\draw (-1.75,1.8) node {{$a_{i_0}$}};
\draw[color=black,ultra thick, ->] (0,0)--(1.5,1.5); 
\draw[color=black,ultra thick, ->] (0,0)--(-1.5,1.5); 
\draw[color=black,ultra thick, ->] (0,0)--(0,-1.5);

\draw (0.,-3.) node {{\footnotesize {\bf Figure 5}: Scheme for a generic
    quantum experiment involving}};
\draw (-0.18, -3.25) node {\footnotesize three sets of random
  outcomes. The outcomes on the three };
\draw (-1.32, -3.5) node {\footnotesize devices are space-like separated
events};
\draw (1.75,1.8) node {{$b_{j_0}$}};
\draw (2.3,1) node {{ $D_B$ }}; 
\draw (-1, -1.8) node {{$D_C$}};
\draw (-2.3, 1) node {{ $D_A$ }};
\draw (0.75,1.1) node  [rotate=45] {{ $\mathscr{S}_B$ }};
\draw (-0.75, 1.1) node [rotate = 315] {{ $\mathscr{S}_A$ }};
\draw (0.2,-0.8) node [rotate = 90] {$\mathscr{S}_C$};
\draw (0.3,0.) node {{$\mathcal{T}$}};
\end{scope}
\end{tikzpicture}
\end{center}}

\newcommand{\figi}{\begin{center}
\begin{tikzpicture}
\begin{scope}[scale=1.2]
\draw[color=black,ultra thin] (-2,-2)--(1.5,1.5); 
\draw[color=black,ultra thin] (2,-2)--(-1.5,1.5); 
\draw[color=black, ->] (-2.5,-2)--(-2.5,0.25); 
\draw[color=black,ultra thick] (2.25,1.75)--(1.75,1.25)--(1.25,1.75)--(1.75,2.25)--(2.25,1.75)--(1.75,1.25);
\draw[color=black,thick] (1.75,1.4)--(1.75,2.1);
\draw[color=black,densely dashed] (2,2)--(2.5,2.5);
\draw[color=black,densely dashed] (1.5,2)--(1.0,2.5);
\draw[color=black,ultra thick] (-2.25,1.75)--(-1.75,1.25)--(-1.25,1.75)--(-1.75,2.25)--(-2.25,1.75)--(-1.75,1.25);
\draw[color=black,thick] (-1.75,1.4)--(-1.75,2.1);
\draw[color=black,densely dashed] (-1.5,2)--(-1,2.5);
\draw[color=black,densely dashed] (-2,2)--(-2.5,2.5);
\draw[color=black,ultra thick, ->] (0,0)--(1.5,1.5); 
\draw[color=black,ultra thick, ->] (0,0)--(-1.5,1.5); 

\draw[color=black,thick] (0,-0.5)--(0,0.5); 

\draw (0.,-3.5) node {{\footnotesize {\bf Figure 2}: Scheme for an
    experiment involving maximally}};
\draw (-2.2, -3.75) node {\footnotesize entangled photons};
\draw (0.75,1) node  [rotate=45] {{\scriptsize Photon}};
\draw (-1.25,-1.04) node  [rotate=45] {{\scriptsize Lightcone}};
\draw (-2.7,-1) node  [rotate=90] {{\footnotesize Time}};
\draw (0.,-0.7) node {M'};
\draw (2.5,2.7) node {{\footnotesize $b_t$}};
\draw (-2.5,2.7) node {{\footnotesize $a_t$}};
\draw (1,2.7) node {{\footnotesize $b_r$}};
\draw (-1,2.7) node {{\footnotesize $a_r$}};
\draw (2,0.8) node {{\scriptsize\parbox{2cm}{{\scriptsize $P_B$ set at angle $\beta$}}}};
\draw (-2,0.8) node {{\scriptsize\parbox{2cm}{ $P_A$ set at angle $\alpha$}}};
\end{scope}
\end{tikzpicture}
\end{center}}

\newcommand{\figii}{\begin{center}
\begin{tikzpicture}
\begin{scope}[scale=1.2]
\draw[color=black,ultra thin] (-2.5,-2.5)--(1.5,1.5); 
\draw[color=black,ultra thin] (1.5,-1.5)--(-2.5,2.5); 
\draw[color=black, ->] (-2.5,-1.5)--(-2.5,1.5); 
\draw[color=black,ultra thick] (2.25,1.75)--(1.75,1.25)--(1.25,1.75)--(1.75,2.25)--(2.25,1.75)--(1.75,1.25);
\draw[color=black,thick] (1.75,1.4)--(1.75,2.1);

\draw[color=black,densely dashed] (2,2)--(2.5,2.5);
\draw[color=black,densely dashed] (1.5,2)--(1.0,2.5);

\draw[color=black,ultra thick] (2.25,-1.75)--(1.75,-1.25)--(1.25,-1.75)--(1.75,-2.25)--(2.25,-1.75)--(1.75,-1.25);
\draw[color=black,thick] (1.75,-1.4)--(1.75,-2.1);
\draw[color=black,densely dashed] (2,-2)--(2.5,-2.5);
\draw[color=black,densely dashed] (1,-2.5)--(1.5,-2);
\draw (1,-2.7) node {{\footnotesize ${a_r}$}}; 
\draw (2.5,-2.7) node {{\footnotesize $a_t$}};
\draw (2.5,2.7) node {{\footnotesize $b_t$}};
\draw (1,2.7) node {{\footnotesize $b_r$}};

\draw[color=black,ultra thick, ->] (0,0)--(1.5,1.5); 

\draw[color=black,thick] (0,-0.5)--(0,0.5); 

\draw[color=black,ultra thick, ->] (1.5,-1.5)--(0,0); 

\draw (0.,-4.1) node {{\footnotesize {\bf Figure 1}: Scheme for an
    experiment involving a single polarized }};
\draw (-3.2,-4.35) node {\footnotesize photon}; 

\draw (1,-0.77) node  [rotate=315] {{\scriptsize Photon}};
\draw (-1.25,-1.04) node  [rotate=45] {{\scriptsize Lightcone}};
\draw (-2.7,0) node  [rotate=90] {{\footnotesize Time}};
\draw (0.,-0.6) node {{\scriptsize Mirror}};

\draw (0.25,1.75) node {{\scriptsize\parbox{2cm}{{\scriptsize $P_B$ at angle $\beta$}}}};
\draw (0.25,-1.75) node {{\scriptsize\parbox{2cm}{$P_A$ at angle $\alpha$}}};

\end{scope}

\end{tikzpicture}
\end{center}}

\newcommand{\figiii}{\begin{center}
\begin{tikzpicture}
\begin{scope}[scale=1.2]
\draw[color=black,ultra thick] (2.25,1.75)--(1.75,1.25)--(1.25,1.75)--(1.75,2.25)--(2.25,1.75)--(1.75,1.25);
\draw (1.75,1.8) node {{$b_{j_0}$}};

\draw[color=black,ultra thick] (2.25,-1.75)--(1.75,-1.25)--(1.25,-1.75)--(1.75,-2.25)--(2.25,-1.75)--(1.75,-1.25);
\draw (1.75,-1.8) node { $a_{i_0}$};
\draw (1.75,-2.7) node {{ $D_A = \{a_i\}_{i \in X}$}}; 
\draw (1.75,2.7) node {{ $D_B = \{b_j\}_{j\in Y} $}};

\draw[color=black,ultra thick, ->] (0,0)--(1.5,1.5); 

\draw[color=black,thick] (0,-0.5)--(0,0.5); 

\draw[color=black,ultra thick, ->] (1.5,-1.5)--(0,0); 

\draw (0.,-4.1) node {{\footnotesize {\bf Figure 3}: Scheme for a generic
    quantum experiment involving}};
\draw (-0.2,-4.35) node {\footnotesize two sets of random
  outcomes. The outcome on device D$_A$};
\draw (-1.5, -4.6) node {\footnotesize causes the outcome on device D$_B$.};

\draw (1,-0.65) node  [rotate=315] {{$\mathscr{S}_A$}};
\draw (1, 0.65) node [rotate=45] {{$\mathscr{S}_B$}};
\draw (0.,-0.7) node {{\scriptsize $\mathcal{T}$}};

\end{scope}

\end{tikzpicture}
\end{center}}

\newcommand{\figiv}{\begin{center}
\begin{tikzpicture}
\begin{scope}[scale=1.2]
\draw[color=black,ultra thick] (2.25,1.75)--(1.75,1.25)--(1.25,1.75)--(1.75,2.25)--(2.25,1.75)--(1.75,1.25);
\draw (1.75,1.8) node {{$b_{j_0}$}};

\draw[color=black,ultra thick] (2.25,-1.75)--(1.75,-1.25)--(1.25,-1.75)--(1.75,-2.25)--(2.25,-1.75)--(1.75,-1.25);
\draw (1.75,-1.8) node { $a_{i_0}$};
\draw (1.75,-2.7) node {{ $D_A = \{a_i\}_{i \in X}$}}; 
\draw (1.75,2.7) node {{ $D_B = \{b_j\}_{j\in Y} $}};

\draw[color=black,ultra thick, ->] (0,0)--(1.5,1.5) ; 

\draw[color=black,thick] (0,-0.5)--(0,0.5); 

\draw[color=black,ultra thick, ->] (0,0)--(1.5,-1.5); 

\draw (0.,-4.1) node {{\footnotesize {\bf Figure 4}: Scheme for a generic
    quantum experiment involving}};
\draw (-0.,-4.35) node {\footnotesize two sets of random
  outcomes. The outcome on device D$_A$ and};
\draw (-1.15, -4.6) node {\footnotesize the outcome on D$_B$ are space-like events.};
\draw (1,-0.65) node  [rotate=315] {{$\mathscr{S}_A$}};
\draw (1, 0.65) node [rotate=45] {{$\mathscr{S}_B$}};
\draw (0.,-0.7) node {{\scriptsize $\mathcal{T}$}};

\end{scope}

\end{tikzpicture}
\end{center}}

\begin{document}
\title{Principle of Relativity for Quantum Theory}
\author{Marco Zaopo}\email{marco.zaopo@unipv.it} 
\affiliation{Dipartimento di Fisica, Universit\`a
  di Pavia, via Bassi 6, 27100 Pavia, Italy}
\email{marco.zaopo@unipv.it}
\date{\today}
\begin{abstract}
In non relativistic physics it is assumed that both chronological
ordering and causal ordering of events (telling whether there exists a
causal relationship between two events or not) are absolute, observer
independent properties. In relativistic physics on the other hand
chronological ordering depends on the observer who
assigns space-time coordinates to physical events and
only causal ordering is regarded as an observer independent
property. In this paper it is shown that quantum theory can be considered as
a physical theory in which causal (as well as chronological) ordering of
probabilistic events happening in experiments may be regarded
as an observer dependent property. We then argue that this result has connections with the
problem of dark energy in cosmology. 
\end{abstract}

\maketitle
\section {Introduction}\label{intro}

The most notable attempts in
formulating a theory that unifies quantum theory and general
relativity are String Theory and Loop Quantum Gravity \cite{polch,rovelli}. The
lack of experiments that could verify or falsify any of the
predictions of the two theories leaves physicists with the
consciousness that something is missing in our current understanding
of nature at the fundamental level. Despite the formulation of both theories
mentioned above depart from very reasonable starting points, they
remain naive about giving a foundational principle to explain the
mathematical formalism of quantum theory. This means that they retain
superposition principle, non-locality and all the counterintuitive
features manifested by quantum theory as natural facts and do not try to
give a motivation for them. This attitude is perhaps justified by the fact
that quantum theory is extremely successful in making predictions.
Until now, no experimental situation has been found in
which the predictions of quantum theory are not satisfied. Such an
extraordinary predicting power has led many physicists to think
that it is not necessary to have a physical intuition
of what is going on at atomic and subatomic scales, it is sufficient to
have a model that can predict whatever can occur in an experiment.     
This pragmatic attitude would be the right one if theoretical physics
accomodated all phenomena experienced in nature in a unique and
coherent model. Despite the many successes of the Standard Model and
the potentiality of string theory and loop quantum gravity, there is large consensus among
physicists that we are far away to have such a unified picture. 
This has recently led physicists to turn the attention back to the
problem of foundations quantum theory with a new slant given by the
emergence of quantum information \cite{axioh,axiob,axiom,axioq,mia,fivel,goyal}. In this paper it is
analyzed the mathematical structure of quantum
theory (as used in the field of
quantum information) from a novel point of view enlighting the
interplay between quantum features and causal structure of space-time events. In quantum
theory events correspond to
probabilistic outcomes and the only predictable and verifiable
statements regard correlations between outcomes happening on different
devices located in distinct regions of space. Since these outcomes are
also thought to be events happening in space-time, it is always
assumed an absolute causal ordering for them. A set of physical events $\mathcal{E}$, like
those that can happen in a quantum experiment, possesses a causal
ordering if, for these events, it is defined a \emph{causal
  structure}. This means that for any pair of events, $\chi_a, \chi_b \in
\mathcal{E}$, one of the following must hold:
\begin{itemize}
\item $\chi_a$ causes $\chi_b$
\item $\chi_b$ causes $\chi_a$
\item $\chi_a$ does not cause $\chi_b$ and $\chi_b$ does not cause
  $\chi_a$ (they are space-like events)
\end{itemize}
For example, $\chi_a$
could be a preparation contained in a preparations ensemble for a
quantum system of a certain type while
$\chi_b$ could be an outcome of a measurement caused by that preparation. In
this case $\chi_a$ causes $\chi_b$. $\chi_a$ and $\chi_b$ could also be two outcomes obtained
respectively in two measurements performed in parallel on a bipartite state of a
composite system. In this case $\chi_a$ and $\chi_b$ are indeed two
space-like separated events. The main result of this paper is that, in
quantum theory, any experimental situation of the former type
mentioned above can be
considered as equivalent to a situation of the latter type.
This equivalence is such that the two experiments can be interpreted as
the same experiment viewed by two different observers that make two
different assumptions regarding the causal ordering of
events happening in the experiment. To prove this it is shown in section \ref{csqt} that, in a generic quantum experiment involving two sets of random outcomes
happening on distinct devices, the mathematical expression of the
joint probability of any two outcomes calculated by one observer, can
be mapped, by means of a simple transformation rule, into the expression for the joint probability of the same
two outcomes calculated by another observer that assumes a different
causal ordering of events with respect to the first. After having
generalized this concept to experiments involving more sets of random
outcomes we are led to
introduce a new physical principle, the "Principle of Relativity
of Causal Structure", and to put it as a foundational
principle for quantum theory. From this principle we understand that
a possible way to move towards a theory of quantum gravity is to retain causal
structure of physical events as an observer dependent
property. Here we take a first step in this direction comparing the idea that causal structure is an
observer dependent property with the role causal structure plays in
general relativity (see section \ref{csgr}). It is
argued that the situation in general relativity theory is somewhat
opposite to the one outlined in quantum theory. If, in quantum
theory, causal ordering of probabilistic events can be regarded as an
observer dependent property, this clearly cannot hold in general
relativity. In general relativity, the causal
ordering of two events is represented by the value of the metric
function evaluated at the two space-time points representing those
events. Einstein's equations relate the metric function to the
stress-energy tensor representing energy density in the portion of
universe including the two events. This implies that, in general
relativity, whether it exists a causal influence between two events or
not, ultimately depends on energy density that is an objective,
physically measureable quantity and hence cannot be regarded as an
observer dependent property. 
Elevating the principle of relativity of causal structure to
universal principle
finally leads us to consider dark energy not as a conceptual problem but
as an essential ingredient of our current understanding of the
universe (see section \ref{pode}).

This research is important for two reasons. The first is that it gives
a new foundational principle to motivate the mathematical structure of
quantum theory. The second is that, in doing this, it is possible to
argue that one of the most puzzling features of modern theoretical
physics, dark energy, could be explained elevating the above
foundational principle for quantum theory to a universal
principle. Clearly this would imply that Einstein's theory of general
relativity should be definitely abandoned and should be
elaborated a deeper theory of the cosmos to explain observational data.

\section{Space-time and causal structure}

A space-time is, roughly speaking, a mathematical representation of physical events. For any set of
physical events $\mathcal{E}$, given two events $p,q \in
\mathcal{E}$ one of the three mutually exclusive alternatives must
hold:
\begin{itemize}
\item $p$ is the cause of $q$

\item $q$ is the cause of $p$

\item $p$ is not the cause of $q$ and viceversa. 
\end{itemize}  

Specifying one of the three alternatives for every pair of events
leads to define the \emph{causal structure} of the set
$\mathcal{E}$. The first of the above alternatives means that $q$ is
in the future of $p$ while the second means that $p$ is in
the future of $q$. This, in turn, is equivalent to say that it exists a physical system
that is present in correspondence with both events $p$ and $q$. The
third indeed means that it is impossible for a
physical system to be present in correspondence with both events $p$
and $q$ (i.e. $p$ and $q$ are causally independent).


In non relativistic (or newtonian) space-time, given an event $p$ for all other events $q$
it must hold one of the following alternatives: (i) $q$ is in the
future of $p$; (ii) $q$ in the past of $p$ (iii) $q$ happens at
the same time of (is simultaneous with) $p$. Regarding this latter case, the events simultaneous with
$p$ constitute points of a three dimensional euclidean space. This
distinction comes from the fact that, in non relativistic space-time, the chronological ordering of
events is the same as their causal ordering. If $p$ and $q$ are one the
cause of the other then necessarily one must happen before the other
while if $p$ and $q$ are causally independent then they must
necessarily happen at the same time.

In relativistic space-time the latter fact above mentioned does not hold anymore. In particular, two causally independent
events can be simultaneous for some observers and have a different
chronological ordering for another observer. From this fact the
set of events $q \in \mathcal{E}$ that constitutes the past and future of $p$ are
represented respectively as points of a four dimensional cone while the
set of events that are not in past nor in the future of $p$ are
represented by points outside those two cones embedded in euclidean four
dimensional space.

Both in non relativistic and relativistic
physics, two different observers can in principle assign different coordinates
to a physical event $p$ because they move relatively to one another. In newtonian
space-time if observer $O$ labels $p$ with coordinates $(t,x,y,z)$ and
$O'$ moves with velocity $v$ in the $x$ direction passing $O$ at
$t=x=y=0$ then the coordinate labels assigned to $p$ by $O'$ are
$t'=t, x'=x-vt, y'=y, z'=z$. In special relativity, i.e. if $v$ is
sufficiently close to the speed of light $c$, those relations
become $t' = (t-vx/c^2)/(1-v^2/c^2)^{1/2}), x'=
(x-vt)/(1-v^2/c^2)^{1/2}), y'=y,z'=z$. Since two
different observers looking at the same physical process
must describe the same physics independently of
their state of motion relative to one another, it is clear that the
above transformations of coordinates leave unaffected any
significant physical property. This implies that coordinate labels
do not have any intrinsic physical significance since they
only depend on which observer labels physical events.

The causal structure of any set of events $\mathcal{E}$
is incorporated in any space-time that can be
used to represent those events.
Moreover, it constitutes an absolute, observer independent property, contrary to
the space-time coordinates assigned to them. For this reason, in both
newtonian and relativistic space-time there
exist specific quantities represented by functions of the coordinates of any two points
$p$ and $q$, that remain
unchanged in changing point of view from one observer to another. In newtonian physics
this function is the time interval $\Delta_t= t_{p}-t_q$. In special relativity
this function is $M = - (\Delta t)^2 + 1/c^2[(\Delta x)^2 + (\Delta y)^2
+ (\Delta z)^2)]$. In general relativity this function is
represented by the metric tensor associated to a manifold
representing a solution of Einstein's equations. The value of these
functions evaluated at every pair of points $(p,q)$ encodes the causal
structure of events. 

We can thus say that both newtonian and relativistic space-time are
different mathematical ways to model a set of events with an
absolutely (i.e. independently of observers) defined causal structure.

Outcomes happening on devices in quantum experiments are supposed to
be events in space-time. From this fact they possess a definite,
observer independent causal structure. In the next section we
are going to show that, although an absolute causal structure of events is a
background assumption in the usual formulation of quantum theory,
the quantum formalism permits to compute
correlations for events happening in experiments in such a way that
their causal structure can be regarded as an observer dependent property.

\section{Causal structure in quantum theory}\label{csqt}

In what follows we are going to show that causal structure in quantum
theory may be regarded as an observer dependent property rather than
fixed in an absolute way. 

\subsection{Experiments involving two sets of random outcomes}

Consider the quantum experiment involving a polarized photon
shown in figure 1.

\figii

We have two polarizers $P_{A}$ and $P_{B}$, the
former aligned at an angle $\alpha$ and the latter aligned at an angle
$\beta$. A photon passes first through $P_{A}$ is reflected by a
mirror and then passes through $P_{B}$. For the experiment to take
place the photon must either be transmitted or be reflected by
polarizer $P_{A}$. Hence associated to $P_{A}$ we have two possible
mutually exclusive outcomes that we indicate $\{a_{r}, a_t\}$. After
the mirror reflection the photon enters $P_{B}$ and then is absorbed
by some photon counter. In order to be counted the photon must either
be transmitted or be reflected by $P_{B}$. Hence also associated to
$P_{B}$ we have two mutually exclusive outcomes that we call $\{b_r,
b_t\}$. The information contained in the experiment is
represented by the joint probability distribution $p(a_i,b_j)$ with $(a_i,b_j)
\in \{a_{r},a_t\} \times \{b_r,b_t\}$. The arrows linking the various devices represent the path
followed by the photon. In particular the arrow pointing out of
$P_A$ means that the photon is an output system for polarizer
$P_A$. The arrow pointing inside $P_B$ means that the photon is an
input system for $P_B$. The lightcone and the arrow of time are drawn
to remark that two events associated to any pair of outcomes
$(a_i,b_j)$ are one the cause of the other. Indeed
there is a physical system, i.e. the photon, that carries the
information regarding the probability distribution $\{p(a_i)\}_{a_i
\in \{a_r,a_t\}}$ from $P_{A}$ to $P_{B}$. This means that if
the probability distribution $\{p(a_i)\}_{a_i\in \{a_r,a_t\}}$ changes
and becomes $\{q(a_i)\}_{a_i\in \{a_r,a_t\}}$ then also the probability
distribution $\{p'(b_j)\}_{b_j \in \{b_r,b_t\}}$ changes.  
The above discussion implies that any pair of outcomes $(a_i,b_j)$ is
such that $a_i$ causes $b_j$ and the correlations between the sets of
random outcomes $\{a_i\}$ and $\{b_j\}$ are due to a causal influence.

Consider now the experiment shown in figure 2.

\figi

We have the same
polarizers $P_{A}$ and $P_{B}$ involved in the previous
experiment and for simplicity we assumed they are aligned in the same
direction as before. Two photons in an entangled state of zero total angular momentum
start from a source of entangled photons, $M'$, and reach independently
$P_{A}$ and $P_{B}$ respectively. After they have passed the
polarizers they are absorbed by two photon counters placed after
$P_{A}$ and $P_{B}$ respectively. For the experiment to take
place, both the photons must be either transmitted or reflected by the
respective polarizers before being detected. Hence also in this case,
associated to both $P_A$ and $P_B$, there are two sets of mutually exclusive
outcomes $\{a_r,a_t\}$ and $\{b_r,b_t\}$ and these represent the same
outcomes as in the previous experiment. The joint probability distribution $p(a_i,b_j)$ with $(a_i,b_j)
\in \{a_{r},a_t\} \times \{b_r,b_t\}$ contains the information about
the experiment. In figure 2 there are two
arrows pointing inside polarizers $P_A$ and $P_B$ respectively. Also
in this case are drawn the lightcone and the arrow of time to help
visualizing that any pair of
outcomes $(a_i,b_j) \in \{a_r,a_t\} \times \{b_r,b_t\}$ represents two
space-like events. 

The two experiments described above seem very
different. The latter involves, for each repetition of the experiment, a pair of entangled photons
while the former involves a single photon. This difference in
their physical description is due to the fact that in each run of the
experiment, it is assumed in one case that the pair of outcomes $(a_i,b_j)$ are
one the cause of the other (the casual relationship being represented
by a photon travelling from $P_A$ to $P_B$) and in the other case that they are two
space-like events (since they are due to two causally independent systems). We can thus say that the main difference in the two
above experiments relies on how, each run of the experiment, the
outcomes $(a_i,b_j) \in \{a_i\}_{i=r,t} \times \{b_j\}_{j=r,t}$ are embedded in space-time. The setup in figure 1
involves three devices, the two polarizers $P_A$ and $P_B$ and a
mirror $M$. The experiment in figure 2 also involves three devices,
two of them are the same polarizers as before while the third device, $M'$
is a source of entangled photons. For
the experiment in figure 1 the photon is an output system
for $P_A$, it is an input and an output for $M$ while it is an
input system for $P_B$. For the experiment in figure 2 the photons involved
may be regarded as two outputs for $M'$ and as two input systems for $P_A$
and $P_B$ respectively. Hence the difference between
the two experiments is that a photon is seen as an
output system for $P_A$ (and in consequence as an input for $M$) in the
experiment of figure 1 while it is seen as an input system for $P_A$ (and
in consequence as an output for $M'$) in the experiment of figure
2. From the above discussion we can say that the existence of a causal relationship between
the region where lies $P_A$ (where happen outcomes $\{a_r,a_t\}$) and
the region where lies $P_B$ (where
happen $\{b_r,b_t\}$) is equivalent to assign a specific input/output
structure for the devices involved in the experiment. We can thus say
that the input/output structure of the devices involved in the
experiment is equivalent to the causal structure assigned to the
outcomes associated to those devices.

In both the situations described above it is assumed a definite causal
structure between the region of space where lies $P_A$
and that where lies $P_B$. This means that it is assumed in an
absolute way either that between
region $P_A$ and region $P_B$ there exists a causal
relationship or that regions $P_A$ and $P_B$ are space-like separated. 
On the other hand, every experiment in
quantum theory is intrinsically probabilistic and whatever an observer
might experience reduces to correlations between outcomes happening on
two devices in distinct regions. This observation
suggests that a definite causal structure between region $P_A$ and
region $P_B$ could not be significant in predicting joint
probabilities for events happening in these two regions. Since
correlations between events is the only
observable and physically predictable property in quantum theory, it
could be the case that the two experiments described in figure 1 and 2
are simply a different way to describe the same experiment. 
Indeed they both define a joint probability
distribution between the values of the same
pair of observables (polarizations along $\alpha$ and $\beta$),
they refer to the same type of system (the photon) and differ only
because in the former it is assumed a causal relationship between regions $P_A$ and $P_B$
while in the latter it is assumed that
regions $P_A$ and $P_B$ are space-like separated. In what follows we
will show that the mathematical formalism of quantum theory is
consistent with the above suggestion.

Suppose that an experimenter sets up one of the two experiments
illustrated above, say the one in figure 1 for definiteness. Two
observers look at this experiment without knowing the nature of device $M$
and the actual input/output structure between the devices.
The observers experience the correlations between the set of outcomes
$\{a_r,a_t\}$ associated to $P_A$ and the set
$\{b_r,b_t\}$ associated to $P_B$. To one observer
it is said that $M$ is a mirror and that the setup is actually
the one in figure 1. To the other observer it is indeed said that
$M$ constitutes a source of maximally entangled photons and that the setup
corresponds to the one in figure 2. We will call the former observer
$O_1$ and the latter observer $O_2$. Comparing figure 1 and 2 we can
readily understand that $O_1$ assumes that photons constitute outputs for $P_A$
and inputs for $P_B$ while $O_2$ assumes that photons constitute
inputs for both $P_A$ and $P_B$. These two assumptions cannot be
verified (or falsified) by the two observers experiencing correlations between
$\{a_r,a_t\}$ and $\{b_r, b_t\}$. Hence they can calculate the joint
probability distribution $\{p(a_i,b_j)\}$ with $(a_i,b_j) \in
\{a_r,a_t\} \times \{b_r, b_t\}$ on the base of the information they respectively
have regarding causal structure. We will now show that for all
$(a_i,b_j) \in \{a_r,a_t\} \times \{b_r,b_t\}$, the probability calculations of
observers $O_1$ and $O_2$, although apparently different, reduce to
the same calculation and give rise to the same probability
value. According to this we may conclude that the two experiments in
figures 1,2 are the same experiment seen by two different observers who
assume a different causal structure between the regions where are
situated polarizers $P_A$ and $P_B$. 

$O_1$ assumes that the polarizer $P_A$ prepares an ensemble
represented by $p |a_r\rangle\langle a_r| + (1-p) |a_t\rangle\langle
a_t|$. By now, let us assume $p = 1/2$ for simplicity. The probability of seeing
outcome $b_t$ in correspondence of $P_B$ given that it is prepared a
photon in state $a_r$ is $
p(b_t|a_r) = |\langle b_s|a_r \rangle|^2
$
thus the joint probability is:
\begin{equation}\label{opt1}
p_{O_1}(a_r,b_t) = 1/2 \langle b_t|a_r \rangle^2
\end{equation}

$O_2$ indeed assumes that $M$ is a source of entangled photons in
state $|\psi\rangle = 1/\sqrt{2} (|a_r a_r \rangle + |a_t a_t \rangle)$. The joint
probability of seeing outcomes $a_r$ and $b_t$ calculated by $O_2$ is:
\begin{equation}\label{opt2}
p(a_r,b_t) = | \langle a_r| \otimes \langle b_t| 1/\sqrt{2} (|a_r a_r \rangle + |a_t a_t \rangle)|^2
\end{equation} 
But the above equation actually reduces to (\ref{opt1}). Expliciting
(\ref{opt2}) we have:
\begin{equation}\label{eqo12}
\begin{aligned}
p_{O_2}(a_r,b_s) = 1/2 (\langle a_r|a_r \rangle^2 \langle b_t |a_r \rangle^2
+ \langle a_r|a_t\rangle^2 \langle b_t|a_t\rangle^2   + \\
+2 \langle a_r | a_r \rangle \langle b_t |a_r\rangle \langle a_r|a_t
\rangle \langle b_t |a_t \rangle) \;\;\;\;\;\;\;\;\;\;\;\;\;
\end{aligned} 
\end{equation}
and all terms in (\ref{eqo12}) are zero except the first thus we can write: 
\begin{equation}\label{eqo12'}
p_{O_2}(a_r,b_t) = 1/2 \langle b_t|a_r \rangle^2
\end{equation}
Clearly the above reasoning is true for every pair $(a_i, b_j) \in
\{a_r,a_t\}\times\{b_r,b_t\}$. Moreover it is simple to convince
ourselves that nothing would change if we assumed that the set
up prepared by the experimenter at which $O_1$ and $O_2$ both look was that in figure 2 in place of the
one in figure 1. This simple example shows that the assumptions of $O_1$ and $O_2$
regarding causal structure of regions $P_A$ and $P_B$ are equivalent for the purpose of calculating joint
probabilities. Whatever an observer of anyone of the above experiments can
experience are correlations between outcomes in region $P_A$ and
outcomes in region $P_B$, and whatever he can predict are joint
probabilities for the outcomes in those regions. Hence, the fact that between those two regions
there exists a causal relationship or not is a property that depends on the assumption of an
observer and cannot be fixed absolutely for all observers in any way.

Note that the equivalence stated above derives from he fact that
(\ref{opt2}) is an alternative way of writing (\ref{opt1}). If it were
not so then causal structure could not be an observer dependent
property. Indeed the correlations between region $P_A$ and region $P_B$
depend on the probability distribution $\{p(a_i,b_j)\}$ and if the
probability distribution calculated by observer $O_2$ was different
from that calculated by observer $O_1$ then one of the observers,
$O_2$, would predict wrong probabilities and would become aware, after camparing
his calculations with those of $O_1$, that correlations are effectively
due to a causal relationship between $P_A$ and $P_B$. This implies
that the equivalence of the two above situations is a consequence of
how in quantum theory are performed probability calculations for the
experiments illustrated in figure 1 and 2.


The two situations considered above are far from being the most general
experiments correlating random outcomes in two regions of space. The equivalence of
(\ref{opt1}) and (\ref{opt2}) could infact be a numerical coincidence. In the
remaining part of this section we will prove that the above
property is a general feature of quantum theory. 
We will consider a generic quantum experiment in which two
devices D$_A$ and D$_B$ display two sets of random outcomes
$\{a_i\}_{i\in X}$ and $\{b_j\}_{j\in Y}$ respectively with $X$ and
$Y$ two sets of outcomes. The information on such
correlations is contained in the joint probability distribution
$\{p(a_i,b_j)\}_{(i,j)\in X\times Y}$. As in the previous example, we suppose that two observers $O_1$ and $O_2$ are looking
at the experiment; $O_1$ assumes that correlations between D$_A$
and D$_B$ are due to a system causally correlating the outcomes in
$\{a_i\}_{i\in X}$ to those in  $\{b_j\}_{j\in Y}$ while $O_2$ assumes
that D$_A$ and D$_B$ lie in space-like separated regions. 


 

\vspace{0.2 cm}

{\bf{Observer $O_1$}}

$O_1$ assumes that
correlations are due to a causal relationship.
In this case a system
$\mathscr{S}$ carries the information of the probability
distribution of one of the sets of outcomes, say $\{a_i\}_{i\in X} $
with probability distribution $\{p_i\}_{i \in X}$, from device D$_A$ to device
D$_B$. The experiment seen by $O_1$ is
represented in figure 3.

\figiii

System $\mathscr{S}_A$ is the
output system for
D$_A$ while $\mathscr{S}_B$ is the input system for D$_B$. Of course they may be
the same system and we distinsuish them only for the purpose of
distinguishing the arrow associated to D$_A$ from that associated to
D$_B$ in the above diagram. An outcome
$a_{i_0}\in \{a_i\}_{i \in X}$ is a
preparation belonging to the preparations ensemble $\{a_i\}_{i\in X}$ with
associated probability ditribution $\{p_i\}_{i\in X}$. The ensemble is represented by a density
matrix $\rho$ and a POVM $\{\bf{a_i}\}_{i \in X}$ as follows:
\begin{equation}\label{ro}
\rho = \sum_{i\in A} \text{Tr}[\bf{a_i}\rho] \frac{\sqrt{\rho} \;\bf{a_i} \sqrt{\rho}}{\text{Tr}[\bf{a_i}\rho]}
\end{equation}
To achieve as
much generality as we can, we will not make any restriction on $\rho$
a part from assuming that it does not represent a pure state since
otherwise the
outcomes on device D$_A$ would not be random anymore contrary to our
initial assumptions.
The ensemble $\rho$
causes probabilistically an outcome $b_{j_0} \in \{b_j\}_{j \in Y}$ on
device D$_B$. In the most general case, this is represented
by an element of a POVM
$\{\bf{b_j}\}_{j\in Y}$ for hilbert space $\mathcal{H}_{\mathscr{S}_B}$. The ensemble
represented by $\rho$ before causing outcome $b_{j_0}$  will eventually
undergo an evolution that is generically represented
by a Completely Positive Trace Preserving (CPTP)
map $\mathscr{T}$. Its Kraus decomposition is $\sum_{m} K^m\otimes
K^{m\dagger} $ with $K^m = \sum_{ef} K^m_{ef} |e\rangle_{B}
{}_{A}\langle f|$ Kraus operator \cite{kraus} ($\{|e\rangle\}_{e=1}^{d_B},
\{|f\rangle\}_{f=1}^{d_A}$ are orthonormal basis for hilbert space $\mathcal{H}_{\mathscr{S}_B}$ and $\mathcal{H}_{\mathscr{S}_A}$ respectively). 
We now explicit the evolution of ensemble $\rho$ by means of
transformation $\mathscr{T}$. The density matrix
obtained after the evolution is:
\begin{equation}\label{tr}
\mathscr{T}({\rho}) = \sum_{m,ef,cd} K_{ef}^mK_{cd}^{m*}
|e\rangle_B {}_A\langle f|\rho |c\rangle_A{}_B\langle d|
\end{equation}
Using the fact
that $\sum_{m} K^m\otimes
K^{m\dagger} $ can be written as:
\begin{equation} 
\sum_{m,ef,cd} K_{ef}^mK_{cd}^{m*}
|c\rangle_A {}_A\langle f| \otimes |e\rangle_B{}_B\langle
d|\end{equation} and the polar decomposition of $\rho$ we have:
\begin{equation}\label{tro}
\mathscr{T}({\rho}) = \text{Tr}_A[\sum_{m,ef,cd} K_{ef}^mK_{cd}^{m*}
\sqrt{\rho}|c\rangle_A {}_A\langle f|\sqrt{\rho} \otimes
|e\rangle_B{}_B\langle d|]
\end{equation}
Note that, for the polar decomposition of $\rho$ to be uniquely
defined, one must assume $\rho$ to be full rank in
$\mathcal{H}_{\mathscr{S}_A}$. The density matrix obtained after the evolution can thus be
written as $\mathscr{T}({\rho}) = \text{Tr}_A[\mathscr{T}_{\rho} ]$
where we define:
\begin{equation}\label{tr}
\mathscr{T}_{\rho} : = \sqrt{\rho} \otimes I_B [\sum_{m} (K^m \otimes
K^{m\dagger})] \sqrt{\rho} \otimes I_B
\end{equation}
where $I_B$ is the identity matrix on $\mathcal{H}_{\mathscr{S}_B}$. From
(\ref{tr}) we see that
the evolution of ensemble $\rho$ can be
represented as an operator acting on
$\mathcal{H}_{\mathscr{S}_A} \otimes \mathcal{H}_{\mathscr{S}_B}$. 
The probability calcualted by observer 1 is then:
\begin{equation}\label{oalfa'}
p_{1}(a_{i_0},b_{j_0}) = \text{Tr}_B[\bf{b_{j_0}} \text{Tr}_A[\mathscr{T}_{\rho} \bf{a}_{i_0}]]
\end{equation}  

\vspace{0.2 cm}

{\bf{Observer $O_2$}}

$O_2$ indeed assumes
that correlations are not due to a causal relationship. This means that the two sets of outcomes
constitute two measurements performed in parallel on two copies of system
$\mathscr{S}$. In figure 4 it is represented the same experiment
of figure 3 as seen by observer $O_2$ assuming that the
regions in which are situated $D_A$ and $D_B$ are space-like
separated.

\figiv

$\mathscr{S}_A, \mathscr{S}_B$ constitute now
two causally independent inputs for devices D$_A$ and D$_B$. The two systems
are both outputs of a common source denoted as
$\tau$ in the above figure. This can be represented by a
bipartite state $\tau_{AB}$ that
permits the observer to calculate the joint probability $p(a_{i_0}, b_{j_0})$ for all
pairs of outcomes as follows:
\begin{equation}
p_{2}(a_{i_0},b_{j_0}) = \text{Tr}_{AB} [\bf{a_{i_0}}' \otimes \bf{b_{j_0}}' \tau_{AB}]
\end{equation} 
where $\bf{a_{i_0}}'$ and $\bf{b_{j_0}}'$ are elements of the POVMs
$\{\bf{a_i}'\}_{i\in X}$, $\{\bf{b_j}'\}_{j\in Y}$ corresponding
respectively to outcomes $a_{i_0}$,
$b_{j_0}$.
\vspace{0.2 cm}

{\bf{Assumptions of observers $O_1$ and $O_2$ are equivalent}}

We are now going to prove the following statement:
Given the mathematical objects used to
calculate joint probabilities of the outcomes by $O_1$, there exists a unique choice of
mathematical objects that permits $O_2$ to calculate the same joint
probability distribution of outcomes. 
Before proving the above statement we recall the discussion regarding
the equivalence between input/output structure and causal structure in
quantum experiments. The only difference between the experiment seen
by $O_1$ and the experiment seen by $O_2$ is that
$\mathscr{S}_A$ is assumed as an output for D$_A$ by $O_1$ while is
assumed as input for D$_A$ by $O_2$. This becomes apparent comparing
figure 3 with figure 4. Based on this observation, we now give the rule that permits to
prove the statement done at the beginning of this paragraph.
\begin{quote}\emph{\bf{Transformation Rule:}} If a system $\mathscr{S}$,
  with hilbert space $\mathcal{H}_{\mathscr{S}}$ is
an input (output) for $O_1$ and an output (input)
for $O_2$, then the operators involving
$\mathcal{H}_{\mathscr{S}}$ used by $O_1$ are the transposed on
$\mathcal{H}_{\mathscr{S}}$ of those used by $O_2$. 
\end{quote}
From the above rule, if $\bf{a_{i_0}}$ represents an element of the preparation ensemble
$\rho$ of $O_1$, $\bf{a_{i_0}^T}$ represents the corresponding
measurement outcome for $O_2$.  For the same reason, the bipartite state
$\tau_{AB}$ has the following expression:
\begin{equation}\label{t12}
\tau_{AB} =  \mathscr{T}_{\rho}^{T_A} = \sqrt{\rho}^T \otimes I_B [\sum_{m} (K^m \otimes
K^{m\dagger})^{T_A} ]\sqrt{\rho}^T \otimes I_2  
\end{equation}
Where ${}^{T_A}$ denotes partial transposition on hilbert space
$\mathcal{H}_{\mathscr{S}_A}$. First we have to prove that (\ref{t12}) is a normalized bipartite
state. This can be seen defining the normalized bipartite state on two copies of $\mathscr{S}_A$, $|\Phi\rangle_{AA'}$:
\begin{equation}\label{fi}
|\Phi\rangle_{AA'} = \sqrt{\rho}^T \otimes I_{A'} 
\sum_j | j \rangle_{A} \otimes |j \rangle_{A'}
\end{equation}
where $\{|j\rangle\}_{j=1}^{d_A}$ is an orthonormal basis for space $\mathcal{H}_{\mathscr{S}_A}$. 
Exploiting (\ref{fi}) we can write:
\begin{equation}\label{isot}
\mathscr{I} \otimes \mathscr{T} (|\Phi\rangle\langle\Phi|) = \tau_{AB}
\end{equation}
where $\mathscr{I}$ is the identity map on $\mathcal{H}_{\mathscr{S}_A}$ and $\mathscr{T}$ represents the evolution defined
above. From (\ref{isot}) we can see that $\tau_{AB}$ is a
normalized bipartite state since $\mathscr{T}$ is a TPCP map acting on
system $\mathscr{S}_A$ and $|\Phi\rangle\langle\Phi|$ is a normalized
bipartite state.
The probability $p_{1}(a_{i_0},b_{j_0})$ expressed in (\ref{oalfa'})
calculated by $O_1$ is then
equal to the probability $p_{2}(a_{i_0},b_{j_0})$ calculated by $O_2$, namely: 
\begin{equation}\label{ogamma}
p_{2}(a_{i_0},b_{j_0}) = \text{Tr}_{AB} [{\bf{a_{i_0}}}^T \otimes {\bf{b_{j_0}}}
\mathscr{T}_{\rho}^{T_A}] = p_{1}(a_{i_0},b_{j_0})
\end{equation}
This expression represents the probability for a given pair of
outcomes $(a_{i_0},b_{j_0}) \in \{a_i,b_j\}_{(i,j) \in X\times Y}$ to jointly happen. 
This proves the statement done at the beginning
of this paragraph. 

In conclusion, every experiment in quantum theory is intrinsically probabilistic and whenever it
correlates two sets of random outcomes displayed by two
devices in two distinct regions of space, an observer
can only experience correlations between these two sets of outcomes
and can only predict their joint probabilities. The causal structure
of these two regions, namely wether the correlations have a causal
origin or not, is always assumed a priori and cannot be subject to a physical
verification. From this fact it follows that if two observers look at one such
experiment and for some reason an observer assumes that correlations are due
to a causal relationship and the other observer assumes that they are not,
they cannot become aware of differences between their respective probabilistic predictions
and the joint probabilities originated by the experiment. 

\subsection{Experiments involving more sets of random outcomes}

Generalizing the result obtained above to experiments
involving more than two sets of outcomes presents some
subtleties. Consider an experiment
involving three sets of random outcomes appearing in three distinct regions
of space, say regions A,B,C, such that the outcomes in A cause
the outcomes in B and these in turns cause the outcomes in C. Let us
suppose that the random outcomes happening in A,B,C are
$\{a_i\}$, $\{b_j\}$, $\{c_k\}$ respectively. A physical system
$\mathscr{S}$ passing through the three regions constitutes the causal
influence propagating from A to B and then from B to C. From an
operational point of view $\mathscr{S}$ is the output of region A, the
input and the output of region B and the input of region C. An outcome
in region B thus represents a possible evolution of $\mathscr{S}$. In
quantum theory a system evolution is represented by a CPTP map and is a
deterministic notion. The only way to take into account randomness in
region B is thus to consider convex combinations of CP maps that
decrease the trace of states. An observer assuming an input/output structure of regions A,B,C modified with
respect to the one given above, does never arrive to assume A,B,C
as three space-like separated regions. Conversely, an experiment where A,B,C
are three space-like separated regions and in which the outcomes in the
three regions are correlated, is due to a tripartite entangled
state. An observer assuming, for this experiment, a different
input/output structure, can never arrive to assume that A,B,C are such
that outcomes in A cause outcomes in B and that these in turns cause
outcomes in C. From these examples we see that when we take into
account three regions of space A, B, C, displaying correlated random
outcomes, if an observer is able to calculate joint probabilities of
the outcomes assuming these three regions as
space-like separated, there cannot exist an observer assuming that
oucomes on A cause outcomes on B that in turns cause
outcomes on C.          
In order to generalize the result in the previous section to experiments involving more
than two sets of random outcomes we thus simply consider that different observers of
the same experiment can in principle assume a different input/output
structure for the devices involved. 
Suppose now to have an experiment in which there are three devices, D$_A$, D$_B$,
D$_C$ in regions A,B,C respectively displaying random correlated outcomes and that an observer
$O_2$, in order to predict the joint probabilities of the outcomes,
assumes that A,B,C are three space-like separated regions. Let the set of outcomes
on the three devices be $\{a_i\}_{i \in X} \times \{b_j\}_{j \in Y} \times \{c_k\}_{k \in Z}$ and the
associated joint probability distribution be
$\{p(a_i,b_j,c_k)\}_{i,j,k \in X \times Y \times Z}$. Let
$\mathscr{S}_A, \mathscr{S}_B, \mathscr{S}_C$ be
the systems to which the outcomes on D$_A$, D$_B$, D$_C$, refer respectively. $O_2$ assumes that $\mathscr{S}_A$,
$\mathscr{S}_B$, $\mathscr{S}_C$ are respectively three inputs for
devices D$_A$, D$_B$ and D$_C$. This is represented in figure 5

\figv

Another observer, $O_1$, assumes that systems $\mathscr{S}_A$ and
$\mathscr{S}_B$ are inputs for D$_A$ and D$_B$ respectively and system
$\mathscr{S}_C$ is an output for D$_C$. This is represented in figure 6.

\figvi

It is easy to see
that this situation is not different from the one analyzed in the
above sections. $O_1$ assumes the outcomes on devices D$_C$ as
representing preparations belonging to some preparation ensemble
represented by a density matrix $\rho$:
\begin{equation}
\rho = \sum_{k \in Z} \text{Tr}[\bf{c_k}\rho] \frac{\sqrt{\rho} \;\bf{c_k} \sqrt{\rho}}{\text{Tr}[\bf{c_k}\rho]}
\end{equation}
Moreover he assumes that outcomes on devices D$_A$ and D$_B$ are POVMs
$\{{\bf{a_i}}\}_{i\in X}$
$\{{\bf{b_j}}\}_{j\in Y}$. The ensemble $\rho$ undergoes an
evolution represented by a CPTP map $\mathscr{T}$ with Kraus
decomposition $\sum_{m} K^m\otimes
K^{m\dagger} $ resulting in a density matrix $\mathscr{T}(\rho)$
having the following expression:
\begin{equation}
\mathscr{T}(\rho) = \text{Tr}_C[\mathscr{T}_{\rho}]
\end{equation}
where 
\begin{equation}\label{tro1}
\mathscr{T}_{\rho} = \sum_{m,ef,cd} K_{ef}^mK_{cd}^{m*}
\sqrt{\rho}|c\rangle_C {}_C\langle f|\sqrt{\rho} \otimes
|e\rangle_{AB}{}_{AB}\langle d|
\end{equation}
We see that the only difference between (\ref{tro1}) and (\ref{tro}) is
that one of the hilbert spaces considered in (\ref{tro1}) explicitly
refers to the hilbert space of a composite system $\mathcal{H}_{\mathscr{S}_{AB}}$. From the
transformation rule stated in the previous section, $O_2$
assumes that outcomes on devices D$_A$, D$_B$ and D$_C$
are respectively represened by the POVMs $\{{\bf{a_i}}\}_{i\in X}$,
$\{{\bf{b_j}}\}_{j\in Y}$, $\{{\bf{c_k}^T}\}_{k\in Z}$ where ${}^{{T}}$
denotes transposition. The three devices
seen by $O_2$ are indeed correlated by a tripartite entangled
state $\tau_{ABC}$ that, according to the transformation rule of the
previous section, is written as:
\begin{equation}
\tau_{ABC} = \mathscr{T}_{\rho}^{T_C}
\end{equation}
$O_1$ and $O_2$ experience the same joint probability
distribution since:
\begin{equation}
\text{Tr}_{ABC} [\tau_{ABC} {\bf{a_{i_0}}} \otimes {\bf{b_{j_0}}} \otimes
{\bf{c_{k_0}^T}}] = \text{Tr}_{AB} [{\bf{a_{i_0}}} \otimes {\bf{b_{j_0}}}
\text{Tr}_C[\mathscr{T}_{\rho} {\bf{c_{k_0}}}]]
\end{equation}
In the same way they can be treated all the cases in which different
observers assume different input/output labels for
$\mathscr{S}_A$, $\mathscr{S}_B$ and $\mathscr{S}_C$. Based on these
arguments it can be seen that analogous results hold for generic
experiments in which an arbitrary
number of devices display correlated random outcomes.

\subsection{Related work}\label{qcs}

The work presented here has connections with three
other works by Hardy
\cite{hardy3}, Oreshkov-Costa-Bruckner \cite{costa} and Leifer-Spekkens \cite{Leif1}.
All these works present formulations of
quantum theory in which calculations of joint probabilities for
sets of outcomes in distinct regions of space can be performed with a
mathematical formalism that is not sensitive of the causal structure
imposed to the regions. The mathematical objects that permit this to
be done are called Causaloid, Process Matrix and Quantum Conditional
State for the three works cited above respectively. Note that quantum theory, as is currently regarded, is a formalism that \emph{is} sensitive to what
causal structure is imposed to different correlated
regions. For two devices in two regions of space displaying correlated random outcomes such that the outcomes
on one device cause those on the other, we have the following
mathematical representation: one set of outcomes is represented by a density
matrix for a single system that is subject to some evolution represented by a linear map;
the other set of oucomes is represented by a set of positive
operators that sum to the identity. For two devices displaying correlated random outcomes in
two space-like separated regions we have indeed the following
mathematical representation: the two sets of outcomes are represented
by two sets of positive operators that sum to the identity; a state
for the composite system, represented by
a density matrix for this system, originates the correlations between the outcomes.
On the other hand the analysis done in this paper suggests that
this may not be the proper way to approach the theory. Indeed,
investigating more deeply quantum theory from this point of view we
have shown that the two above mathematical representations are more
similar than one could expect.

Here we enlight similarities and analogies of this paper with
\cite{hardy3,costa,Leif1}. The operator
defined as $\mathscr{T}_{\rho}$ in (\ref{tr}), i.e. the evolution by means of map
$\mathscr{T}$ of ensemble $\rho$, has a lot of analogies
with a process matrix \cite{costa}. Indeed they both represent a way to
calculate joint probabilities for outcomes happening in different
regions of space that is insensitive to what causal structure is
assumed for the regions. This is because the operator $\tau_{AB}$
establishing correlations for outcomes in space-like separated
regions is a mathematical object of the same nature of
$\mathscr{T}_{\rho}$ (being simply its partial transposition). The main difference between the
situation depicted in the previous section and the process matrix
formalism is that in the former case, outcomes are represented by POVM
elements while in the latter
case they are represented by quantum operations. Hence we could regard
$\mathscr{T}_{\rho}$ as a process matrix for POVM elements. There are
even more strict similarities with the work in \cite{Leif1}. To see
this note that $\tau_{AB}$ in (\ref{t12}) is simply the joint state
obtained with an acausal conditional state (see equation (9) in \cite{Leif1}).
$\mathscr{T}_{\rho}$ in (\ref{tr}), on the other hand is a causal
joint state, i.e. the joint state obtained with a causal conditional
state (see section IIIE in \cite{Leif1}). Relationships of the
work in \cite{hardy3} with the work presented here (as with the other two
works) are less explicit. The work in \cite{hardy3} has the remarkable
feature of being formulated in a general probabilistic
framework. To achieve such generality it becomes necessarily
more abstract and the formulation of quantum theory in this
framework suffers of such abstractness. The main idea of the
causaloid is that embedding probabilistic physical
processes in space-time (hence giving to probabilistic events a causal
structure) is an instance
of compression of information. The starting point to reach this
conclusion is that causal structure and space-time in physics may not be regarded as something
really existing in an objective way. Indeed this is very close to the
starting point we adopted in the previous section and to the idea that
causal structure of probabilistic outcomes is an observer dependent
property.

\subsection{Relativity of causal structure and no-signalling}

In this subsection we will discuss the no-signalling principle in light of the result
obtained so far. Although it could seem at first
sight that our result contradicts no-signalling, we will show that
the principle of relativity of causal structure is indeed
consistent with it. In particular we are going to show that there is
no contradiction in relativity of causal ordering of probabilistic
outcomes in a quantum experiment even when it is established in an
absolute way that the correlations among those outcomes are either "signalling" or "no-signalling".

No-signalling is the name of a condition usually formulated in the context of
foundations of quantum theory for outcomes correlations between two space-like
separated devices due to an
entangled state. This condition is elevated to a principle because
it ensures that non-local correlations in quantum mechanics do not
allow istantaneous signalling between two parties. The following
provides the definition of no-signalling condition for outcomes
correlations of two space-like separated devices:

\begin{definition}\label{nosig} \emph{No-signalling condition}

Suppose to have two generic sets of outcomes happening respectively on two devices $D_{A}$ and
$D_{B}$ due to an entangled state $\tau$ of a composite system. We say that
the outcomes correlations obey the no-signalling condition iff the following are both
satisfied:
\begin{itemize}
\item for all $b_{j_0}
  \in \{b_j\}_{j \in Y}$, where $\{b_j\}_{j \in Y}$ is any set of
  outcomes on $D_B$, it holds:
\begin{equation}\label{ns}
p(b_{j_0}|\tau) = \sum_{a_i} p(a_i,b_{j_0}|\tau) = \sum_{a'_i} p(a_i',b_{j_0}|\tau) 
\end{equation}  
for all possible different pairs of sets of outcomes $\{a_{i}\}_{i\in X},
\{a_{i'}\}_{i'\in X'}$ happening on $D_A$.
\item for all $a_{i_0}
  \in \{a_i\}_{i \in X}$, where $\{a_i\}_{i \in X}$ is any set of
  outcomes on $D_A$, it holds:
\begin{equation}\label{ns}
p(a_{i_0}|\tau) = \sum_{b_j} p(a_{i_0},b_{j}|\tau) = \sum_{b_j'} p(a_{i_0},b_{j'}|\tau) 
\end{equation}  
for all possible different pairs of sets of outcomes $\{b_{j}\}_{j\in Y},
\{b_{j'}\}_{j'\in Y'}$ happening on $D_B$.
\end{itemize}
\end{definition}

In \cite{popescu} it is showed that it is possible to formulate models in which this
condition holds and where the outcomes correlations are stronger than
those originated by maximally entangled states. This implies that
no-signalling condition alone cannot be put as a foundational
constraint for quantum correlations. 
In \cite{masanosig} they are explored the consequences of assuming this condition
alone, for a generic non-local probabilistic theory.

The meaning of the above condition is the following. If the
probability distribution and the outcomes are changed on device
$D_A$ ($D_B$) then the probability distribution of the outcomes on
$D_B$ ($D_A$) is not affected. This is the case since the probability
of any outcome $a_{i_0}$ ($b_{j_0}$) that happens on $D_A$ ($D_B$) is
the joint probability of $a_{i_0}$ ($b_{j_0}$) with the outcome
corresponding to the coarse graining of all the outcomes that can
possibly happen on $D_B$ ($D_A$). No-signalling is thus
guaranteed by the fact the correlations are such that changing
something in the statistics of one of the two devices cannot result in
any statistical change in the outcomes on the other device. Such changes, if possible,
would be due to an istantaneous influence since outcomes correlations
are istantaneous, and this would permit istantaneous signalling from
one device to the other.

The apparent tension between no-signalling and relativity of causal
ordering is due to the fact that no-signalling conditions are
formulated using joint probabilities of outcomes and thus provide absolute
statements about the (im)possibility of influencing probability
distributions on one device manipulating a different and space-like
separated device. Provided that causal ordering of probabilistic
outcomes is not an absolute property one could in fact imagine the
following misleading scenario. An observer assumes that two sets of correlated
outcomes are such that the outomes in one set cause those in the other
set; in this case correlations
can be signalling and one of the two conditions in definition \ref{nosig}
can be violated. A second observer assuming
the same outcomes to be space-like separated would then experience
signalling correlations. But this would mean that from his point of view there
could be an istantaneous influence from one device to another. In what
follows we are going to show that this situation is never attained.
To see it is so suppose that an observer $O_1$ is looking to an experiment in
which they are displayed correlated random outcomes on two devices,
$D_A$ and $D_B$ and that he assumes that outcomes on device $D_{A}$
cause those of device $D_B$. Accordingly he assumes the specific
input/output structure depicted in figure 3. From his point of view
the correlations between the outcomes on $D_A$ and those on $D_B$ can
be signalling. This means that $O_1$ can devise situations in which:
\begin{equation}\label{sign}
\sum_{i \in X} p(a_i,b_{j_0}) \neq \sum_{i' \in X'} p(a_i',b_{j_0}).
\end{equation}
(\ref{sign}) means that the preparations ensemble $\{a_i\}_{i \in
X}$ with probability distribution $\{p_i\}_{i\in X}$ represented by a
density matrix $\rho$ is changed into the preparation ensemble
$\{a_i'\}_{i'\in X'}$ with probability distribution $\{p_i'\}_{i'\in
  X'}$ represented by a density matrix $\rho'$ different from $\rho$. An observer $O_2$ on
the other hand assumes that the two devices are indeed space-like
separated thus assuming the input/output structure of figure 4. From
his point of view (\ref{sign}) is by no means
paradoxical. Indeed if we write the explicit expression for (\ref{sign})
as calculated by $O_1$ we have, using (\ref{oalfa'}):
\begin{equation}\label{signtr}
\sum_{i \in X} \text{Tr}_{AB}[{\bf{b_j}} \otimes {\bf{a}_i}
\mathscr{T}_{\rho}] \neq \sum_{i \in X} \text{Tr}_{AB}[{\bf{b_j}} \otimes {\bf{a}_i'}
\mathscr{T}_{\rho'}] 
\end{equation}  
where $\mathscr{T}_{\rho}$ represents the evolution by means of TPCP
map $\mathscr{T}$ of ensemble $\rho$ and similarly for
$\mathscr{T}_{\rho'}$. According to observer $O_2$, following the
transformation rule stated in section \ref{csqt}, (\ref{signtr}) is
rewritten as follows:
\begin{equation}
\sum_{i \in X} \text{Tr}_{AB}[{\bf{b_j}} \otimes {\bf{a}_i^{T}}
\mathscr{T}_{\rho}^{T_A}] \neq \sum_{i \in X} \text{Tr}_{AB}[{\bf{b_j}} \otimes {\bf{a'}_i^{T}}
\mathscr{T}_{\rho'}^{T_A}] 
\end{equation}
This expression means
that the bipartite state $\tau = \mathscr{T}_{\rho}^{T_A}$ has changed
into the bipartite state $\tau' = \mathscr{T}_{\rho'}^{T_A}$ and the
verification of the above inequality may not be ascribed to an
instantaneous influence between two space-like separated devices but
simply to a change in the bipartite state correlating the two devices.

On the other hand, for observer $O_2$ assuming that correlations of
the outcomes on $D_A$ and $D_B$ are due to a bipartite state $\tau$ we
must have that definition \ref{nosig} holds. However, there is nothing
that prevents observer $O_1$ to assume that
outcomes on $D_A$ cause the outcomes on $D_B$. This is the case since
no signalling conditions are only a set of necessary conditions that the
outcomes correlations of two devices must satisfy if the devices are
space-like separated. Indeed, from the Choi-Jamiolkowsky
isomorphism \cite{choi}, \cite{jamio}, observer $O_1$ can always interpret every
bipartite state $\tau$ as the evolution by means of a CPTP map
$\mathscr{T}$ of an ensemble $\rho$ thus assuming no-signalling
correlations between two sets of outcomes as due to a system carrying a causal influence
from one device to another.

\subsection{Two principles for quantum theory}

The work presented here, compared to those reviewed above, has, in our opinion,
a deeper foundational value since it poses
a new physical principle, the observer dependence of causal structure,
as a foundational principle for quantum
theory. This is achieved recognizing the equivalence of input/output
structure and causal structure and showing that the mathematical
formalism of quantum theory is consistent with the assumption that
input/output structure is an observer dependent property. 

We can thus summarize the work done in this section saying that
quantum theory is consistent with the two following principles:

\begin{quote} {\bf Principle of causality}
The input/output structure of the devices involved in a quantum experiment defines the
causal structure of the outcomes happening on those devices.
\end{quote}

\begin{quote}{\bf Principle of relativity of causal structure}
Two observers looking at a given quantum experiment and assuming a
different causal structure for the outcomes involved in the experiment
cannot become aware of differences in their respective probabilistic predictions.
\end{quote}

In the next section, the principle of relativity of causal structure 
will be compared with the role causal structure
plays in general relativity. In particular it is argued that the
situation in general relativity is somewhat opposite to the one
outlined above. This is the case since, in general relativity, whether two events in
two distinct regions of universe are space-like or not is determined
by the metric that, in turn, is related to the stress energy tensor
via Einstein's equations. This implies that in general relativity
causal structure depends on a (in principle) measurable physical quantity,
energy density, and cannot be regarded as an observer dependent property.

\section{Causal structure in general relativity}\label{csgr}

In this section we briefly examine the role causal structure of events
has in general relativity. The main equations of general relativity
are Einstein's equations relating the metric of
a portion of space-time manifold describing a given portion of
universe with the mass/energy content of that portion of
universe. They are often expressed in the following compact form \cite{Wald}: 
\begin{equation}\label{einst}
G_{\mu\nu} = k T_{\mu\nu}
\end{equation}
On the r.h.s. $k$ is a constant and $T_{\mu\nu}$ is the stress-energy
tensor; on the l.h.s $G_{\mu\nu}$ is the Einstein's tensor and has the following
expression:
\begin{equation}\label{einstens}
G_{\mu\nu} = R_{\mu\nu} - \frac{1}{2} R g_{\mu\nu} + \Lambda g_{\mu\nu}
\end{equation}
where $g_{\mu\nu}$ is the metric, $R_{\mu\nu}$ is the Ricci tensor,
$R$ is the Ricci scalar and $\Lambda$ is the cosmological constant. 
On a manifold, ($M$, $g$), a geodesic is a path $x^{\mu}(\lambda)$
characterized by the following equation \cite{Wald, Carroll}:
\begin{equation}\label{geod}
\frac{d^2x^{\mu}}{d \lambda^2} + \Gamma^{\mu}_{\rho\sigma}
\frac{dx^{\rho}}{d\lambda} \frac{dx^{\sigma}}{d\lambda} = 0
\end{equation}
In the above equation  $\Gamma^{\mu}_{\rho\sigma}$ are the
coefficients of the Levi-Civita connection associated to the metric of
the manifold (in general one can use any connection but in general
relativity it is used only the Levi-Civita one). This is written as follows:
\begin{equation}\label{levi}
\Gamma^{\mu}_{\rho\sigma} = \frac{1}{2} g^{\rho\sigma} (\partial_{\mu}g_{\nu\sigma} +
\partial_{\nu}g_{\sigma\mu} - \partial_{\sigma}g_{\mu\nu})
\end{equation}
where $\partial_x$ denotes partial derivative, $g_{xy}$ is the metric
and $g^{xy}$ is its inverse. Equation (\ref{geod}) can be interpreted as the vanishing of the covariant
derivative of $x^{\mu}$ along the path $x^{\mu}(\lambda)$. This means that any vector
on $x^{\mu}(\lambda)$ is transported parallel to itself along this
path. The tangent vector to a
point of the geodesic describes an interval between two points in the
tangent space. If the manifold is a solution of Einstein's Equations, such interval can be time-like, null or space-like
depending on the norm of the vector. Since a geodesic describes a path
along which a tangent vector of the manifold is parallel transported,
we have that if the tangent vector on a given point of the geodesic is
time-like, null or space-like, the tangent vector on any other point
of the geodesic will preserve this property. From this, one interprets
geodesics where the tangent vector is time-like or null as paths
followed respectively by freely falling material particles or
photons. On the other hand if the tangent vector is space-like, then
there is no physical system that can follow the path corresponding to
the geodesics. From this fact we can state that, in general
relativity, given two points in space-time $x_a$, $x_b$,
pertaining to two different regions of universe R$_A$, R$_B$ respectively,
it can exist a
causal relationship between them (namely it is possible for a
material or light particle to start at $x_a$ and cause an effect at
$x_b$) if the two points lie on a time-like or null geodesic. On the other hand it cannot exist a causal
relationship between the two points if they lie on a space-like
geodesic. From (\ref{geod}) and (\ref{levi}) we see that, in last
analysis, the metric tensor is the object characterizing
geodesics. This together with Einstein's equations imply that the stress-energy
tensor representing the energy density in a given portion of universe 
establishes whether between two space-time points it can exist a causal relationship. 

According to general relativity we thus have that the existence (or
non existence) of a causal
relationship between two events depends on the energy density
of the portion of universe in which the events happen and thus on an
objective physical quantity. This means that causal structure in
general relativity should
(in principle) be inferred in an objective
way by whatever observer by means of energy density measurements.  
We used the conditional because it is well known that, on large
cosmological scales, to explain at best observational data it must be introduced dark energy and this poses
various problems from the theoretical point of view. In the following
section we will briefly review these problematic issues. After that we
will discuss the possible relationship that could exist between these
problems and the fact that causal structure in
quantum theory may be regarded as an
observer dependent property.

\section{Dark energy}\label{pode}

In this section we first give a brief review on dark energy.
This material is mostly taken from a review on the
subject done by Carroll \cite{Carroll1}. We then discuss the conclusions reached
in this review in relationship with the observer dependence of causal
structure for outcomes happening in quantum experiments.

\vspace{0.4 cm}

The standard assumption in cosmology is that universe is homogenous
and isotropic. Since in general relativity, universe is described by a
manifold $M$, these
two assumtpions translate into formal statements regarding the geometry
of $M$. Homogeneity means that given two points $p, q$ in $M$ there
exists an isometry that takes $p$ into $q$. Isotropy means that given
a point $p$ in $M$, for any two vectors v and w in $T_p M$, there
exists an isometry such that the pushforward of w under the isometry
is parallel to v. Since the universe is not static, we infer that it
is homogeneous and isostropic in space but not in time. This and the
above assumptions imply that the universe can be foliated in space-like slices such that each
slice is homogeneous and isotropic. Based only on these considerations
it can be shown \cite{Wald, Carroll} that the metric of the universe must have the
following form:
\begin{equation}\label{FRLW}
ds^2 = -dt^2 + a^2(t) d\sigma_3^2(k)
\end{equation}
where $a(t)$ is the scale factor and $d\sigma_{3}^2(k)$ is a metric
for three space which depends on the curvature parameter
$k$. The metric in (\ref{FRLW}) is called the Friedmann Robertson
Lemaitre Walker (FRLW) metric. Note that
Einstein's equations are not taken into account to derive (\ref{FRLW})
since its derivation is based on purely geometrical
arguments. Einstein's equations are used to find the functional form for
$a(t)$. In order to do so it must be made the assumption that matter and energy on large cosmological
scale can be modelled as a perfect fluid and it is choosen an equation
of state relating pressure $p$ to matter and energy density $\rho$ of
the type $p = w \rho$ with $w$ constant. Putting the metric in
(\ref{FRLW}) into Einstein's equations and using the above assumption leads to write Friedman
equations \cite{Wald, Carroll}, i.e. a set of differential equations establishing
the evolution of scale factor in relationship with
curvature, pressure and energy density:
\begin{equation}\label{feq1}
\frac{\ddot{a}}{a} = \frac{4\pi G}{3} (\rho + 3p)
\end{equation}

\begin{equation}\label{feq2}
 (\frac{\dot a}{a})^2 =
  {8\pi G\over 3}\rho - \frac{k}{a^2}
\end{equation}
The quantity on the l.h.s of (\ref{feq2}) is the square of the \emph{Hubble parameter} $H =
\frac{\dot a}{a}$ and can be used to define the value of the \emph{critical density}:
\begin{equation}\label{roc}
\rho_c= \frac{3H^2}{8\pi G}
\end{equation}
The critical density is the value of energy density solving Friedman's
equations for zero spatial curvature, i.e. for a flat universe. 
Exploiting $\rho_c$ one can define the \emph{density parameter} $\Omega =
\frac{\rho}{\rho_c}$ by means of which (\ref{feq2}) can be written as:
\begin{equation}\label{feq2'}
\Omega - 1 = \frac{k}{H^2a^2}
\end{equation}
This shows that whether $k = +1,0,-1$ depends on the magnitude of the
actual (i.e. observed) energy density $\rho$ with respect to critical
density $\rho_c$. If $\Omega<1$ then $k<0$ and the universe
is described by a three dimensional manifold with constant negative
curvature. On the contrary, if $\Omega>1$ then $k>0$ and the universe
is decribed by a three dimensional manifold with constant positive curvature (the
analog in three dimension of a sphere). Finally $\Omega = 1$ implies
$k=0$ and describes a flat universe the associated manifold being
simply a three dimensional euclidean space. 

There are three forms of
energy density usually considered. The first is called \emph{dust} $\rho_d$ and
is composed of non relativistic matter whose pressure is negligible
with respect to its energy density. The second is called
\emph{radiation} $\rho_r$ and is composed of photons and other relativistic particles moving
approximately at the speed of light. The third is \emph{dark energy}
$\rho_{\Lambda}$ coming from the introduction of the cosmological
constant in Einstein's equations. There are strong evidences \cite{Carroll1} that the amount of
total energy density $\rho$ due to dust is negligible with
respect to the amount due to matter ($\rho_m/\rho_d =
10^6$). We thus say that we live in a matter dominated universe and
the relevant contributions to total energy density come from
$\rho_d$ and $\rho_{\Lambda}$.

Observations of the dynamics of galaxies and clusters have shown that
a reasonable value for the density parameter referring to $\rho_d$, is
$\Omega_d = 0.3 \pm 0.1$ \cite{Carroll2}. On the other hand observations of the
anisotropies of the cosmic microwave background are consistent with a
nearly spatially flat universe \cite{Carroll2}. Thus we infer $\Omega \approx 1$. This
implies that the amount of $\rho_{\Lambda}$ to the total energy density is
such that $\Omega_{\Lambda} \approx 0.7$. Measurements of the distance
vs.\ redshift relation for Type Ia supernovae \cite{Carroll3, Carroll4}
have provided evidences that the universe is accelerating i.e.
that $\ddot a > 0$. Since conventional matter could not make the
universe expansion accelarate it is inferred that the component of the
energy density that is responsible for such acceleration is
$\rho_{\Lambda}$. The
most natural candidate component of energy density for
$\rho_{\Lambda}$ is the vacuum energy $\rho_v$. This is
corroborated by the following argument. Let us write (\ref{feq2}) as:
\begin{equation}
  {\dot a}^2 =
  {8\pi G\over 3}a^2\rho - k .
  \label{feq2''}
\end{equation}
If the universe is expanding, then $\rho_d$ must necessarily 
decrease as the particle number density is diluted by expansion, so
$\rho_d \propto a^{-3}$.  
Hence the right-hand side of 
(\ref{feq2''}) will be decreasing in an expanding universe (since
$a^2 \rho$ is decreasing, while $k$ is a constant), 
hence the derivative of $\dot a$ should be negative if one only takes
into account the contribution of $\rho_d$.
The supernova data therefore imply that, to make the
universe accelerate, there must be a source of energy density that
varies more slowly than $a^2 \rho$ i.e. more slowly than $a^{-2}$. Since the
distinguishing feature of vacuum energy is that it is a minimum amount of energy
density in any region, strictly constant throughout spacetime, the
slow variation of $\rho_{\Lambda}$ corroborates the statement that
vacuum energy be the source of
energy density making the expansion of universe accelerate. 
To match the data, it is required a vacuum energy:
\begin{equation}
 \rho_v \approx 
  (10^{-3}{\rm eV})^4 = 10^{-8} {\rm ergs/cm}^3
  \label{rhoobs}
\end{equation}
It is not possible to reliably calculate the expected vacuum energy in
the universe, or even in some specific field theory such as the 
Standard Model of particle physics; at best they can be evaluated 
order-of-magnitude estimates for the contributions from different
sectors. These estimates lead to the following value:
\begin{equation}
\rho_v^{{\rm (theory)}} \sim (10^{27} {\rm ~eV})^4
  = 10^{112}{\rm ~ergs/cm}^3\ .
  \label{rhotheory}
\end{equation}
This value is 120 orders of magnitude (30 if we change units of measurement) greater than the value in
(\ref{rhoobs}). Such a huge discrepancy with observational data
implies that the source of energy density responsible for
the expansion of universe, $\rho_{\Lambda}$, should be something different from the
vacuum energy. This is known as the \emph{cosmological constant problem}. 

As already told the actual model for the universe has 
$\Omega_{\Lambda} = 0.7$ and $\Omega_{d} = 0.3$ but the
relative balance of dark energy and matter changes rapidly as the
universe expands:
\begin{equation}
\frac{\Omega_{\Lambda}}{\Omega_d} = \frac{\rho_{\Lambda}}{\rho_d}
\propto a^{3}
\end{equation}
This is due to the facts pointed out above, namely, that $\rho_{\Lambda}$ should be almost
constant while $\rho_d \propto a^{-3}$. As a consequence, at early times
of the universe's expansion, dark energy was negligible in comparison to
matter and radiation, while at late times matter and radiation are
negligible.  There is only a brief epoch of the universe's history
during which it would be possible to
witness the transition from domination by
one type of component to another. On the other hand, from the fact that $\Omega_{\Lambda}
= 0.7$ and $\Omega_{d} = 0.3$ we conclude that we actually live in
such a transitional period. It seems remarkable that we live during
the short transitional period between those two eras.
The approximate coincidence between matter and dark energies in the
current universe is called the \emph{coincidence problem}.

\vspace{0.4 cm}

Inferring the existence of a source of energy different from
ordinary matter or radiation to explain observational data
in cosmology is not, on its own, a conceptual problem. Problems
arise because it is not possible to explain the origin of this source of
energy in a scenario that is logically consistent with the current
physical knowledge. Thus, the problematic issues of inferring the
existence of dark energy lie in the fact that this inference leads to
logical inconsistencies such as the cosmological constant problem and
the coincidence problem. 

From the above analysis we understood that different methods to
measure the curvature of space-time give rise to different curvature
estimations. Since in general relativity curvature is
related to an objective physical quantity that should have a definite
value, energy density, we have that curvature itself must be uniquely
defined. For the latter fact to be consistent with the former one we
postulate the existence of dark energy (this assumption is
corroborated but not proved by the observations of \cite{Carroll3}, \cite{Carroll4}).

The existence of a uniquely
defined curvature is the consequence of a unique metric
tensor. In general relativity metric tensor is unique because causal structure of
space-time events is assumed to be fixed in an absolute way. This, as
suggested in the previous section, is opposed to what we found in
quantum theory where causal structure of events is relative to an
observer. 

Inferring dark energy is thus directly related to the assumption that
causal structure of space-time events is absolutely defined. However,
if we elevated relativity of causal structure to be a universal
principle, we could not model our universe with a uniquely defined
metric tensor anymore. From this in turn we could conjecture that two different
methods of measuring curvature give rise to different estimations simply
because an absolutely defined causal
structure of space-time events is not physically defineable. This in
turn could eliminate the problem of dark energy at all but would pose
the deeper problem of formulating a theory of the universe completely
different from the one we have at the moment.

\section{Conclusions}

Quantum theory is an extraordinarily successful theory and still lacks
a clear physical explanation. Moreover, the absence of
experiments linking quantum theory with the geometry of space-time leaves
physicists with the consciousness that something is missing in our
current understanding of nature at a fundamental level.
This has renewed efforts in finding foundational principles for
quantum theory in order to find a more general theory.

In this paper it is analyzed the interplay between causal structure
of space-time events and the probabilistic nature of quantum
theory. This analysis leads us to state two principles that can
be put as foundations of quantum theory:  

\begin{quote} {\bf Principle of causality}
The input/output structure of the devices involved in a quantum experiment defines the
causal structure of the outcomes happening on those devices.
\end{quote}

\begin{quote}{\bf Principle of relativity of causal structure}
Two observers looking at a given quantum experiment and assuming a
different causal structure for the outcomes involved in the experiment
cannot become aware of differences in their respective probabilistic predictions.
\end{quote}

Since the only thing that can be predicted and physically verified in
quantum theory are probabilities, the last principle suggests that
causal structure of outcomes happening in quantum experiments is an observer dependent
property. This principle could be a guiding principle to construct a
theory of quantum gravity for the following reason. Quantum theory and
general relativity are both successful and problematic in different and
somewhat opposite aspects. On one hand quantum theory is extremely
successful in making predictions. Until now, no experimental situation has been found in
which the predictions of quantum theory are not satisfied. However there are still difficulties, after
almost 90 years from its birth, to understand its physical meaning. On
the other hand general relativity is not completely satisfactory in making
predictions at large cosmological scales. This is related to the
need to introduce dark energy to explain observational data. General
relativity is, by the way, founded on two extremely clear and intuitive
physical principles, namely, the Einstein's principles of relativity
and equivalence. It is then likely that a theory more
fundamental than the ones we have at the moment will come from a
physical principle that can be put as foundation of quantum theory on
one hand and that can motivate the need to introduce dark
energy to explain observational data at large cosmological scales on
the other hand. The
principle of relativity of causal structure is indeed such a principle
as we discussed in the previous section. If dark energy was not necessary
to explain cosmological observations, and we could estimate the
sources of energy responsible for the inferred dynamics of the
universe, then, in principle, two
observers could not assume different perspectives regarding the
existence of a causal connection between two regions of universe
since this would be absolutely defined by energy density measurements.
Elaborating a theory of quantum gravity starting from the
conclusions of this work is an extremely hard task and
its success is far from being certain. 
The main motivation to try to formulate a new theory according to the
above principle is that, as far as we know, the most plausible
proposal for a source of dark energy is the assumption of a "cosmic aether" permeating all
space whose origin is unknown \cite{ash}. Clearly this cannot be satisfactory since
we are forcing new physical degrees of freedom,
motivated only by the fact that the current model of universe and the
theory underlying it do not properly explain observations.

\end{document}